\documentclass[twocolumn,showpacs,preprintnumbers,amsmath,amssymb]{revtex4}
\usepackage{amssymb}
\usepackage{txfonts}
\usepackage{graphicx}% Include figure files
\usepackage{dcolumn}% Align table columns on decimal point
\usepackage{bm}% bold math
\usepackage{sidecap}
\begin{document}

\title{The effect of dipole-dipole interaction for two atoms with different couplings in non-Markovian environment}% Force line breaks with \\

\author{Yang Li$^{1, 2}$}
\author{Jiang Zhou$^{1, 2}$}
\author{Hong Guo$^{1, 2}$}\thanks{Author to whom correspondence should be
addressed. phone: +86-10-6275-7035, Fax: +86-10-6275-3208, E-mail:
hongguo@pku.edu.cn.} \affiliation{1. CREAM Group, State Key
Laboratory of Advanced  Optical Communication Systems and Networks
(Peking University) and Institute of Quantum Electronics, School of
Electronics Engineering and Computer Science, Peking University,
Beijing 100871, P. R. China\\2. Center for Computational Science and
Engineering (CCSE), Peking University, Beijing 100871, P. R. China
}%

\date{\today}% It is always \today, today,

\begin{abstract}
Using the nonperturbative method, we study the exact entanglement
dynamics of two two-level dipole-dipole interacting atoms coupled to
a common non-Markovian reservoir with different coupling strengths.
Besides analyzing the conditions for the existence of steady state,
we find, though the dipole-dipole interaction could destroy the
stationary entanglement, that relatively strong dipole-dipole
interaction does suppress the disentanglement in the initial period
of time. These results are helpful for the practical engineering of
the entanglement in the future.
\end{abstract}

\pacs{03.67.Mn, 03.65.Yz, 42.50.Pq}

%\keywords{Use showkeys class option if keyword display desired}%
\maketitle

\section{Introduction}
Quantum entanglement is highly relevant to the fundamental issues of
quantum mechanics and plays a central role in the application of
quantum information science. In recent years, the practical schemes
for the creation of quantum entanglement have gained significant
improvements both in theoretical research and in experimental
realization. One of the most popular schemes involves a couple of
spontaneously emitting two-level atoms (qubits) trapped in a
single-mode cavity \cite{DDE1,DDE2,DDE3}, which describes both atoms
or ions trapped in an electromagnetic cavity and circuit-QED setups.
The corresponding investigations reach notable results on the
dynamical evolution of the bipartite entanglement
\cite{DDE4,F1,F2,PM3} and the steady states of the composite
two-atom system \cite{DDE3,DDE4,DDE5,F2,PM3}.

Most of the previous studies base on an essential assumption that
the coupling strengths of the two atoms to the cavity mode are equal
\cite{DDE1,DDE4}, which is a good approximation for the atoms in
microwave cavity. However, in optical cavity, the two atoms usually
no longer possess the equivalent coupling strengths to the cavity
mode \cite{F2,PM3}. In \cite{F2}, the case of different coupling
constant is considered, but the dipole-dipole interaction is
neglected in the very beginning, as in \cite{PM3}.

In this paper, we study the two two-level dipole-dipole interacting
atoms in an optical cavity via a nonperturbative method
\cite{PM1,PM2,PM3}, and focus on the asymptotic and transient
entanglement dynamics of the composite two-atom system.

For stationary entanglement, when the dipole-dipole interaction
exists and the coupling strengths of the two atoms to the cavity
mode are different, the final state of the two-atom system will
become completely separable and the bipartite entanglement will
collapse to zero, which indicates that there is no decoherence-free
subspace in this case \cite{DFS}. This result holds under both the
Makovian and non-Markovian treatment. Thus, we conclude that the
memory effect resulted from the finite correlation time of the
reservoir in non-Markovnian limit does not influence the asymptotic
bipartite entanglement.

It is shown further, for transient entanglement, that strong
dipole-dipole interaction can suppress the decoherence and
disentanglement. However, when the strength of the dipole-dipole
interaction and coupling strength of the atoms to the environment,
which determines the memory effect, are in the same magnitude order,
the dipole-dipole interaction accelerates the decoherence and
disentanglement for some initial states.

Finally, by transforming the true mode Hamiltonian which is used at
the start \cite{PM2,QO} to the quasimode Hamiltonian \cite{PM2}
which separates the memory effect and the damping effect of the
reservoir explicitly, we present a concrete physical interpretation
for all the results derived in this paper.

\section{Theoretical framework}

Consider two two-level atoms with a common zero-temperature bosonic
reservoir, and take their dipole-dipole interaction into account
from the very beginning. Under the rotating wave approximation
(RWA), the Hamiltonian of the composite two-atom system plus the
reservoir is given by ($\hbar=1$) \cite{DDH0,DDH} :
 \[ H = H_0  +
H_{{\mathop{ int}} } + H_{ dd} ,\] with
\begin{equation}
H_0  = \omega _1 \sigma _ + ^{(1)} \sigma _ - ^{(1)}  + \omega _2
\sigma _ + ^{(2)} \sigma _ - ^{(2)}  + \int_{ - \infty }^\infty
{d\omega _k \omega _k b^\dag  (\omega _k )b(\omega _k )}  ,
\end{equation}
\begin{equation}
H_{ int }  = (\alpha _1 \sigma _ + ^{(1)}  + \alpha _2 \sigma _ +
^{(2)} )\int_{ - \infty }^\infty  {d\omega _k g(\omega _k )b (\omega
_k )}  + {\rm h.c.}  ,
\end{equation}
\begin{equation}
H_{ dd}  = K(\sigma _ + ^{(1)} \sigma _ - ^{(2)}  + \sigma _ +
^{(2)} \sigma _ - ^{(1)} ).
\end{equation}

Now, we use the true mode Hamiltonian. Here, $ \sigma _ \pm ^{(j)}$
and $\omega_{j}$ are the inversion operators and transition
frequency of the $j$th qubit ($j = 1, 2$), and $b(\omega_k)$,
$b^\dag(\omega_k)$ are the annihilation and creation operators of
the field mode of the reservoir. The mode index $k$ contains several
variables which are two orthogonal polarization indexes and the
propagation vector $\vec k$. To measure the coupling strength of the
atoms to the cavity mode determined by the atom's relative position
in the cavity, we introduce the dimensionless constant $\alpha_{j}$
\cite{PM3}. The static part of the dipole-dipole interaction is
proportional to $ K = [\vec d \cdot \vec d - 3(\vec d \cdot \vec
r_{12} )(\vec d \cdot \vec r_{12} )/r_{12}^2 ]r_{12}^{ - 3} $, where
$\vec{r}_{12}=\vec{r}_1 - \vec{r}_2$ is the relative position,
$\vec{d}$ is the electric dipole moment of the atom.

For an initial state of the form
\begin{equation}
\left| {\psi (0)} \right\rangle  = (c_{10} \left| e \right\rangle _1
\left| g \right\rangle _2  + c_{20} \left| g \right\rangle _1 \left|
e \right\rangle _2 )\left| 0 \right\rangle _E ,
\end{equation}
since $[H,N] = 0$, where $ N = \int_{ - \infty }^\infty  {d\omega _k
 b^\dag  (\omega _k )b(\omega _k )}  + \sigma _ + ^{(1)}
\sigma _ - ^{(1)}  + \sigma _ + ^{(2)} \sigma _ - ^{(2)}  $, the
time evolution of the total system is confined to the subspace
spanned by the bases $ \{ \left| e \right\rangle _1 \left| g
\right\rangle _2 \left| 0 \right\rangle _E ,\left| g \right\rangle
_1 \left| e \right\rangle _2 \left| 0 \right\rangle _E ,\left| g
\right\rangle _1 \left| g \right\rangle _2 \left| {1_k  }
\right\rangle _E \} $:
\begin{eqnarray}
\label{eq1}
 \left| {\psi (t)} \right\rangle &=&c_1 (t)e^{ - i\omega _0 t} \left| e \right\rangle _1 \left| g \right\rangle _2 \left| 0 \right\rangle _E  + c_2 (t)e^{ - i\omega _0 t} \left| g \right\rangle _1 \left| e \right\rangle _2 \left| 0 \right\rangle _E
  \nonumber\\& &+
\int_{ - \infty }^\infty  {d\omega _k c_{\omega _k } (t)e^{ -
i\omega _k t} \left| g \right\rangle _1 \left| g \right\rangle _2
\left| {1_k  } \right\rangle } _E ,
\end{eqnarray}
where $\left| {1_k } \right\rangle _E $ is the state of the
reservoir with only one exciton in the $k$th mode. Here, we consider
the case in which the two atoms have the same frequency, i.e.,
$\omega_{1}=\omega_{2}=\omega_{0}$. According to the Schr$\rm
\ddot{o}$dinger's equation, the equations for the probability
amplitudes take the form
\begin{equation}\label{eqc10}
i\dot c_1 (t) = \alpha _1 \int_{ - \infty }^\infty  {d\omega _k e^{
- i(\omega _k  - \omega _0 )t} g(\omega _k )} c_{\omega _k } (t) +
Kc_2 (t) ,
\end{equation}
\begin{equation}\label{eqc20}
i\dot c_2 (t) = \alpha _2 \int_{ - \infty }^\infty  {d\omega _k e^{
- i(\omega _k  - \omega _0 )t} g(\omega _k )} c_{\omega _k } (t) +
Kc_1 (t) ,
\end{equation}
\begin{equation}
\label{eqck} i\dot c_{\omega _k } (t) =[{\alpha _1 c_1 (t) + \alpha
_2 c_2 (t)}]g^* (\omega _k )e^{i(\omega _k  - \omega _0 )t} .
\end{equation}

Eliminating the coefficients $c_{\omega _k } (t)$ by integrating Eq.
(\ref{eqck}) and substituting the result into Eqs. (\ref{eqc10}) and
(\ref{eqc20}), we get:
\begin{eqnarray}
\label{eq14} \dot c_1 (t) =  &-& \int_0^t {dt_1 f(t - t_1 )\alpha _1
[\alpha _1 c_1 (t_1 ) + \alpha _2 c_2 (t_1 )]} \nonumber\\&-&
iKc_2(t) ,
\end{eqnarray}
\begin{eqnarray}
\label{eq15} \dot c_2 (t) =  &-& \int_0^t {dt_1 f(t - t_1 )\alpha _2
[\alpha _1 c_1 (t_1 ) + \alpha _2 c_2 (t_1 )]}  \nonumber\\&-& iKc_1
(t) ,
\end{eqnarray}
where the responding function takes the form:
\[f(t - t_1 ) = \int_{ - \infty
}^\infty  {d\omega _k J(\omega _k )} e^{ - i(\omega _k  - \omega _0
)(t - t_1 )}
.
\]

Suppose the atoms interacting resonantly with the reservoir with
Lorentzian spectral density
\[ J(\omega _k ) = \left| {g(\omega _k
)} \right|^2  = {W}^2 \lambda /\pi [(\omega _k  - \omega _0 )^2 +
\lambda ^2 ],
\]
by employing Fourier transform and residue theorem, we get the
explicit form $f(t - t_1 ) = {W}^2 e^{ - \lambda (t - t_1 )}$, where
the quantity $1/\lambda$ is the reservoir correlation time.

In fact, we can obtain the solution by Laplace approach directly.
However, in order to construct a lucid physical insight, we choose
the pseudo-mode approach \cite{PM1,PM2,PM3}. The corresponding
equations are as follows:
\begin{equation}
\label{eqc1}
\dot c_1 (t) = -iW \alpha _1 b(t) - iKc_2 (t)
,
\end{equation}
\begin{equation}
\label{eqc2}
\dot c_2 (t) = -iW \alpha _2 b(t) - iKc_1 (t)
,
\end{equation}
\begin{equation}
\label{eqb}
 \dot b(t) =  - \lambda b(t) - iW
[\alpha _1 c_1 (t_1 ) + \alpha _2 c_2 (t_1 )]
,
\end{equation}
where
\[
b(t) =  - i\int_0^t {dt_1 e^{ - \lambda(t - t_1 )} W[\alpha _1 c_1
(t_1 ) + \alpha _2 c_2 (t_1 )]}
,
\]
Taking Laplace transform,
$c_{1,2}(t)\fallingdotseq\bar{c}_{1,2}(s),\
b(t)\fallingdotseq\bar{b}(s)$, we get the resulting equations
\begin{equation}
s\bar c_1 (s) - c_{10}  =  - iW\alpha _1 \bar b(s) - iK\bar c_2 (s)
,
\end{equation}
\begin{equation}
s\bar c_2 (s) - c_{20}  =  - iW\alpha _2 \bar b(s) - iK\bar c_1 (s)
,
\end{equation}
\begin{equation}
s\bar b(s) - b(0) =  - \lambda \bar b(s) - iW[\alpha _1 \bar c_1 (s)
+ \alpha _2 \bar c_2 (s)]
.
\end{equation}
Further, we introduce vacuum Rabi frequency $R = W(\alpha _1^2  +
\alpha _2^2 )^{1/2}$ and relative coupling strengthes $r_j  = \alpha
_j (\alpha _1^2  + \alpha _2^2 )^{ - 1/2}$ ($j=1,2$). Applying the
inverse Laplace transform, we get the solutions with the initial
condition $b(0)=0$:
\begin{widetext}
\begin{equation}
c_1 (t) = \sum\limits_{s_i } {\mathop {\lim }\limits_{s \to s_i }
\left\{ {(s - s_i )\frac{{c_{10} [s(s + \lambda ) + R^2 r_2^2 ] -
c_{20} [iK(s + \lambda ) + R^{\rm{2}} r_1 r_2 ]}}{{s^2 (s + \lambda
) + R^2 s + K^2 (s + \lambda ) - 2iKR^2 r_1 r_2 }}} \right\}} e^{s_i
t}
,
\end{equation}
\begin{equation}
c_2 (t) = \sum\limits_{s_i } {\mathop {\lim }\limits_{s \to s_i }
\left\{ {(s - s_i )\frac{{c_{20} [s(s + \lambda ) + R^2 r_1^2 ] -
c_{10} [iK(s + \lambda ) + R^2 r_1 r_2 ]}}{{s^2 (s + \lambda ) + R^2
s + K^2 (s + \lambda ) - 2iKR^2 r_1 r_2 }}} \right\}} e^{s_i t}
,
\end{equation}
\begin{equation}
b(t) =  - iR\sum\limits_{s_i } {\mathop {\lim }\limits_{s \to s_i }
\left\{ {(s - s_i )\frac{{(r_1 c_{10}  + r_2 c_{20} )s - iK(r_1
c_{20}  + r_2 c_{10} )}}{{s^2 (s + \lambda ) + R^2 s + K^2 (s +
\lambda ) - 2iKR^2 r_1 r_2 }}} \right\}} e^{s_i t}
.
\end{equation}
where $s_i$ belongs to the roots of the following equation for $s$:
\begin{equation}
\label{Eq3} {s^2 (s + \lambda ) + R^2 s + K^2 (s + \lambda ) -
2iKR^2 r_1 r_2 }=0
.
\end{equation}
\end{widetext}

\textit{Asymptotic analysis}.---In physical problems considered in
this paper, $\bar c_{1,2}(s)$ have poles which lie either on the
imaginary axis, giving an oscillatory term, or in the left-hand half
plane, giving a term which exponentially decays in magnitude.
Consider function $\bar c_{1,2}(s)$ each having a single simple pole
$s_0=i\theta$ lying on the imaginary axis, and all other poles lying
in the left-hand half plane. The residue at $s_0=i\theta$ is
\begin{equation}
\label{finalvalue}
\mathop {\lim }\limits_{s \to i\theta } (s -
i\theta )\bar c_{1,2} (s)\exp (i\theta t) .
\end{equation}
The other residues are of the form ${\mathop {\lim }\limits_{s \to
s_i } h_i \exp (s_i t)}$, where $h_i$ is a constant determined by
$s_i$. Since each $s_i$ has a negative real part, the residue decays
exponentially in magnitude as $t$ increases. So, whether the values
of $c_1$ and $c_2$ are zero as $t \rightarrow \infty$ depends on
wether Eq. (\ref{Eq3}) has a pure imaginary solution. Hence, it is
easy to figure out that the conditions for the existence of the
steady state are
\begin{equation}
\label{criterion} K = 0\  \rm{or}\  \alpha_1 = \alpha_2 ,
\end{equation}
which means unless there is no dipole-dipole interaction or the
coupling strengths of the two atoms to the cavity mode are equal,
$c_1(t)$ and $c_2(t)$ will tend to be zero as $t \rightarrow
\infty$.

In the $ \{ \left| e \right\rangle _1 \left| e \right\rangle _2
,\left| e \right\rangle _1 \left| g \right\rangle _2 ,\left| g
\right\rangle _1 \left| e \right\rangle _2 ,\left| g \right\rangle
_1 \left| g \right\rangle _2 \} $ basis, the reduced density matrix
of the two atoms is given by:
\begin{equation}\label{eq2}
\rho _a (t) = \left( {\begin{array}{*{20}c}
   0 & 0 & 0 & 0  \\
   0 & {\left| {c_1 (t)} \right|^2 } & {c_1 (t)c_2^* (t)} & 0  \\
   0 & {c_2 (t)c_1^* (t)} & {\left| {c_2 (t)} \right|^2 } & 0  \\
   0 & 0 & 0 & {1 - \left| {c_1 (t)} \right|^2  - \left| {c_2 (t)} \right|^2 }  \\
\end{array}} \right)
.\end{equation} We choose the concurrence $C(t)$ \cite{Concurrence}
to measure the entanglement of the two atoms. According to the
definition, the expression of the concurrence of $\rho_a$ is given
by \cite{Concurrence}:
\begin{equation}
C(t) = 2\left| {c_1 (t)c_2^* (t)} \right| = 2\left| {c_1 (t)}
\right|\left| {c_2 (t)} \right|
.
\end{equation}

First, consider $\alpha_1=\alpha_2$, that is, the coupling constants
of the two atoms to the cavity mode are the same. The asymptotic
entanglement evaluated by the concurrence is given by:
\[
C(t \to \infty ) =\left| {c_{10}  - c_{20} } \right|^2/2 .
\]
This indicates that stationary entanglement is only determined by
the initial state of the system.

Next, consider $K=0$, that is, the dipole-dipole interaction is
neglected. The asymptotic entanglement is given by
\[
C(t \to \infty ) = 2\left| {\alpha _R^{ - 1} c_{10}  - c_{20} }
\right|\left| {\alpha _R c_{20}  - c_{10} } \right|/(\alpha _R^2  +
\alpha _R^{ - 2} )
,
\]
where $ \alpha _R  = r_1/r_2$. This indicates that the entanglement
is determined by the initial state and the ratio of the coupling
constants of the two atoms to the cavity mode. This situation has
been discussed in details in \cite{PM3}.

\begin{figure*}
\centering
\begin{minipage}[c]{.55\textwidth}
\includegraphics[width=4.in]{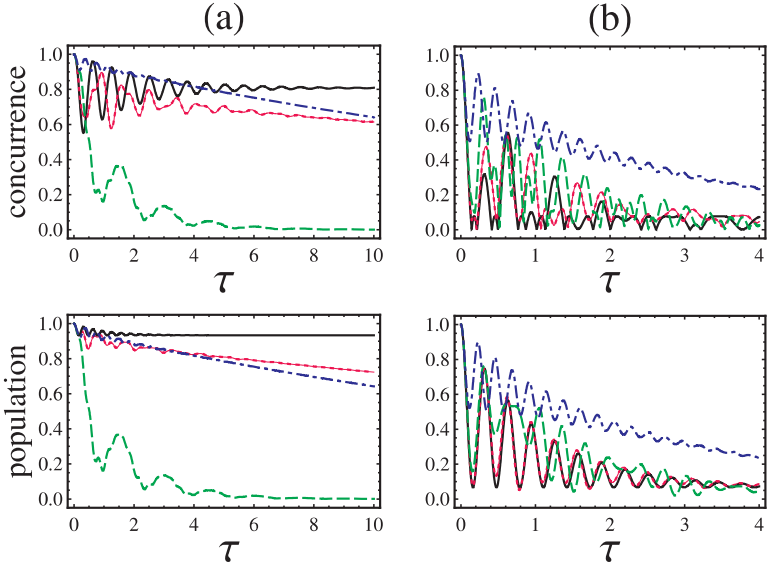}
\end{minipage}%
\hfill
\begin{minipage}[c]{.4\textwidth}
\centering \caption{\small (color online) Time evolution of the
concurrence and population, with the initial state being (a) $\left|
{\varphi _ - } \right\rangle$, and (b) $\left| {\varphi _ + }
\right\rangle$, both with $R = 10$, $\lambda = 1$ and $r_1 =
\sqrt{3}/2$, for the cases of (i) no dipole-diploe interaction
${\cal K}=0$ (solid curve), (ii) weak dipole-dipole interaction
${\cal K}=2$ (short-dashed curve), (iii) intermediate dipole-dipole
interaction ${\cal K}=7$ (long-dashed curve), and (iv) strong
dipole-dipole interaction ${\cal K}=20$ (long-short-dashed
curve).}\label{fig1}
\end{minipage}
\end{figure*}

\textit{Dynamical analysis}.---In this part, we will focus on the
entanglement dynamical evolution of the composite system in a good
cavity, i.e., for $\cal R\rightarrow\infty$ with ${\cal R}=
R/\lambda$. In FIG. \ref{fig1}, we show the concurrence as a
function of $\tau  = \lambda t$ in the good cavity limit. We compare
the dynamics of two maximally entangled states (MES: $\left|
{\varphi_ \pm } \right\rangle $) for four different values of
relative strengths of dipole-dipole interaction, namely, ${\cal K} =
0,2,7,20$, with ${\cal K}=K/\lambda$ which represents the ratio of
the strength of the dipole-dipole interaction to the coupling
strength of the atoms to the cavity field. The explicit expressions
of $\left| {\varphi_ - } \right\rangle $ and $\left| {\varphi _ +}
\right\rangle$ are as follows:
\begin{equation}
\left| {\varphi _ \pm  } \right\rangle  = \frac{1}{{\sqrt 2
}}(\left| e \right\rangle _1 \left| g \right\rangle _2  \pm \left| g
\right\rangle _1 \left| e \right\rangle _2 ).
\end{equation}

We choose the coupling parameter $r_1$ as $\sqrt{3}/2$, which is a
typical constant in our case. Other values of $r_1$ which exclude
$r_1=1/\sqrt{2}$ show qualitatively similar behavior. Here, We note
that when $r_1=r_2=1/\sqrt{2}$, $\left| {\varphi _ - }
\right\rangle$ spans an eigenspace of the Hamiltonian of the
composite system with and without dipole-dipole interaction, which
means that once the initial state is $\left| {\varphi _ - }
\right\rangle$, both the population and concurrence will stay at
their maximal value 1 as time increases irrespective of what the
relative strength $\cal K$ is.

For a MES $\left| {\varphi _ - } \right\rangle$, and for the case of
${\cal K}=0$, the population and concurrence oscillate at the very
beginning, but as time increases the composite two-atom system
collapses to a steady state, which means that the black plots for
both the population and concurrence will converge to a straight line
with a fixed value as $\tau$ increases. However, when we take into
account the dipole-dipole interaction (${\cal K}\neq0$) with
$r_1\neq r_2$, both the population and concurrence perform damped
oscillations and tend to be zero inevitably as time increases which
means that the composite system has no steady state and will lose
their bipartite entanglement completely just as our asymptotic
analysis describes. When ${\cal K}=2$, the entanglement collapses
faster with the growing strength of the dipole-dipole interaction.
On the contrary, when ${\cal K}=20$, the entanglement collapses more
slowly with the increasing strength of the dipole-dipole
interaction. The fastest disentanglement happens when the strength
of dipole-dipole interaction and the coupling strength of the atoms
to the cavity field are in the same order of magnitude (${\cal K}
\approx 7$), given all other conditions the same, as shown in FIG.
\ref{fig1} (a). Whereas, for a MES $\left| {\varphi _ + }
\right\rangle$, the disentanglement becomes slower as the
dipole-dipole interaction become stronger, which means the
dipole-dipole interaction can inhibit decoherence and
disentanglement, as shown in FIG. \ref{fig1} (b). In the practical
engineering of entanglement, this result suggest that if it is hard
to gain equal couplings between the atoms to the cavity, one can
simply choose an appropriate distance to make the couplings of the
atoms to the cavity to be much smaller or bigger than the
dipole-dipole interactions. This will, at least, keep the
entanglement alive for a longer time.

In FIG. \ref{fig1}, we choose $\lambda / W = 0.1$, which can be well
realized within the current experimental condition \cite{ECQED}. A
more intuitive impression of the role of the dipole-dipole
interaction can be get from FIG. \ref{fig2}. The explanations of
these results will be given in details in the following section.

\section{Explanations by quasimode Hamiltonian}
In order to erect a concrete explanation for these phenomena, we
convert the true mode Hamiltonian with Lorentz spectral density into
the quasimode form, because in the truemode Hamiltonian, the
coupling between the atoms and infinite modes makes it hard to
separate the damping effect and the memory effect. Using the method
in \cite{PM2}, we get:
\begin{eqnarray}
H = H_{0}' + H_{int}' + H_{dd} +H_{damping},
\end{eqnarray}
\begin{eqnarray}
H_0 '  &=& \omega _1 \sigma _ + ^{(1)} \sigma _ - ^{(1)}  + \omega
_2 \sigma _ + ^{(2)} \sigma _ - ^{(2)}  + \omega _0 a^\dag  a
\nonumber\\
 & &+ \int_{ - \infty }^\infty  {d\Delta \Delta c^\dag
(\Delta )c(\Delta )}  ,
\end{eqnarray}
\begin{equation}
H_{ int }'  = W[(\alpha _1 \sigma _ + ^{(1)}  + \alpha _2 \sigma _ +
^{(2)} )a + (\alpha _1 \sigma _ - ^{(1)}  + \alpha _2 \sigma _ -
^{(2)} )a^\dag],
\end{equation}
\begin{eqnarray}
H_{ damping } = (\lambda /\pi )^{1/2} \int_{ - \infty }^\infty
{d\Delta [a^\dag c(\Delta ) + ac^\dag  (\Delta )]},
\end{eqnarray}
\noindent where $c^\dag  (\Delta ),c(\Delta)$ are the creation and
annihilation  operators of the continuum quasimode of frequency
$\Delta$, $H_{dd}$ and the other parameters are the same as before.

From the above Hamiltonian, the composite two-atom system only
interacts with one discrete mode and their coupling coefficients are
just the transition strength $\alpha_i W$ ($i = 1, 2$). The discrete
mode interacts with a set of continuum modes \cite{PM2} and their
coupling strength only contains the width of the Lorentzian spectral
density $\lambda$ which is a constant. This means that if we let the
two atoms and the discrete mode be a new system and the continuum
modes be the reservoir, the behavior of the new system is exactly
Markovian. The physical interpretation of the quasimode Hamiltonian
has been discussed in details in \cite{PM2}, and more details about
the relations between the quasimode and the true mode Hamiltonian
are given in the appendix. Now we use the quasimode Hamiltonian to
explain the results derived above.

The first thing is the origin of the Non-Markovian memory effect. As
we know, the memory effect originates from the finite lifetime of
the photon in the cavity \cite{PM2,PM3}. According to Eqs.
(26)-(29), the atoms only interact with the discrete mode, while the
irreversible process only happens in the interaction between the
discrete mode and the continuum modes. So, the reabsorbing
phenomenon that causes the entanglement to oscillate, as shown in
FIG. \ref{fig1}, just results from the interaction between the atoms
and the discrete mode, which is a reversible process. Yet, due to
the coupling between the discrete and the continuum modes, the
photon escapes into the continuum modes and never comes back, for
the process is exactly Markovian. This means that the memory effect
is a finite time phenomenon, which fits the results in FIG.
\ref{fig1} where the oscillating phenomenon is only obvious at the
very beginning.

\begin{figure}[h]
\centering
\includegraphics[width=2.0in]{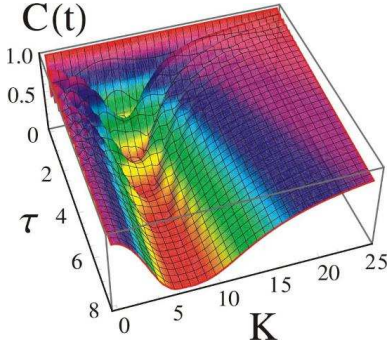}
\caption{\small (color online) Time evolution of concurrence as a
function of dipole-dipole interaction strength $K$, for the initial
state $\left| {\varphi _ - } \right\rangle$ with $R = 10$, $\lambda
= 1$ and $r_1 =\sqrt{3}/2 $.}\label{fig2}
\end{figure}

The second thing is the disentanglement suppressing effect in the
strong dipole-dipole interaction. From Eqs. (26)-(29), we can see
that the atoms are only coupled to the discrete mode, while the
damping process only happens between the discrete and the continuum
modes. This indicates that if the exciton only exists in the atoms,
the damping process does not happen. The damping of the atoms
happens only when the exciton in the atomic system is transferred to
the discrete mode by the interaction between them and then photon
escapes to the continuum modes before it comes back to the atomic
system again. That is to say, the probability of the exciton
transferring from the atom to the discrete mode is related to the
disentanglement. As shown in FIG. \ref{fig2}, if the dipole-dipole
interaction is much stronger than the interaction between the atom
and the discrete quasimode, the disentanglement happens very slowly.
So, in terms of quantum perturbation theory, the dynamics of the
atoms is mainly determined by $H_{dd}$. Because the initial state
$\left| {\varphi _ - } \right\rangle$ form an invariant subspace of
$H_{dd}$, the atoms will stay in this state forever in the zeroth
order perturbation consideration, and the disentanglement is caused
only by the higher-order corrections, which are relatively small
quantities. Therefore, the disentanglement happens slowly. However,
when the two strengths are in the same order of magnitude, $H_{int}$
is not small quantity compared to $H_{dd}$, and the higher-order
correction can not be treated as perturbation, and thus the
disentanglement happens very fast. The analogous discussion goes
well also for the case of the strength of dipole-dipole interaction
$K$ being much smaller than the strength of the interaction between
the atom and the discrete quasimode $W$.

\section{conclusion}
In this paper, we extend the study of the entanglement dynamics to a
realistic situation where the dipole-dipole interaction is not
neglected and the couplings of the atoms to the field are different.
We find, in this situation, that there is no steady state
entanglement. This result is valid for both the Markovian and
non-Markovian environment. Nevertheless, though the dipole-dipole
interaction could destroy the asymptotic entanglement, strong
dipole-dipole interaction does suppress disentanglement effect under
some conditions. These results are helpful for the practical
engineering of the entanglement in the future.

\appendix
\section*{APPENDIX: \ THE RELATIONS BETWEEN THE QUASIMODE HAMILTONIAN AND THE TRUE
MODE HAMILTONIAN}
The transform relation from the true mode
Hamiltonian to the quasimode Hamiltonian is given by:
\begin{eqnarray}
b(\omega _k ) &=& \frac{{(\lambda /\pi )^{1/2} }}{{(\omega  - \omega
_0) + i\lambda }}[a + (\lambda /\pi )^{1/2} {\rm P}\int_{ - \infty
}^\infty  {d\Delta \frac{{c^\dag  (\Delta )}}{{\omega  - \Delta }}}
\nonumber\\
& &+ \frac{{\omega  - \omega _0 }}{{(\lambda /\pi )^{1/2} }}\int_{ -
\infty }^\infty  {d\Delta c^\dag  (\Delta )} ]
.
\end{eqnarray}
The transform relation from the quasimode Hamiltonian to the true
mode Hamiltonian is given by:
\begin{equation}
a = (\lambda /\pi )^{1/2} \int_{ - \infty }^\infty  {d\omega _k
\frac{{b(\omega _k )}}{{\omega _k  - \omega _0  - i\lambda }}}
,
\end{equation}
\begin{eqnarray}
c(\Delta ) &=& (\lambda /\pi ){\rm P}\int_{ - \infty }^\infty
{d\omega _k \frac{{b(\omega _k )}}{{(\omega _k  - \Delta )(\omega _k
- \omega _0  - i\lambda )}}} \nonumber\\
& &+ \int_{ - \infty }^\infty {d\omega _k \frac{{\omega _k  - \omega
_0 }}{{\omega _k  - \omega _0 - i\lambda }}b(\omega )}
,
\end{eqnarray}
Using the quasimode Hamiltonian, the initial state is given by:
$\left| {\psi (0)} \right\rangle {\rm{ = }}c_{10} \left| e
\right\rangle _1 \left| g \right\rangle _2 \left| 0 \right\rangle _a
\mathop  \otimes \limits_\omega  \left| {0_\omega  } \right\rangle -
c_{20} \left| e \right\rangle _1 \left| g \right\rangle _2 \left| 0
\right\rangle _a \mathop  \otimes \limits_\omega  \left| {0_\omega }
\right\rangle $. Since $[N,H] = 0$, where $N = \sigma _ + ^{(1)}
\sigma _ - ^{(1)}  + \sigma _ + ^{(2)} \sigma _ - ^{(2)}  + a^\dag a
+ \int_{ - \infty }^\infty  {d\Delta c^\dag  (\Delta )c(\Delta )} $,
we can write the state of the total system as
\begin{eqnarray}
 \left| { \psi (t)} \right\rangle   &=& c_1 (t)\left| e \right\rangle _1 \left| g \right\rangle _2 \left| 0 \right\rangle _a \mathop  \otimes \limits_\omega  \left| {0_\omega  } \right\rangle  + c_2 (t)\left| g \right\rangle _1 \left| e \right\rangle _2 \left| 0 \right\rangle _a \mathop  \otimes \limits_\omega  \left| {0_\omega  } \right\rangle  \nonumber\\
  &+& b(t)\left| g \right\rangle _1 \left| g \right\rangle _2 \left| 1 \right\rangle _a \mathop  \otimes \limits_\omega  \left| {0_\omega  }
  \right\rangle \nonumber\\
  &+& \int_{ - \infty }^\infty  {d\omega c(\omega )\left| g \right\rangle _1 \left| g \right\rangle _2 \left| 0 \right\rangle _a \left| {1_\omega  } \right\rangle }
.
\end{eqnarray}
According to Schr$\rm \ddot{o}$dinger's equation, we find that the
equations for $c_1 (t)$, $c_2 (t)$, and $b (t)$ are just the same as
Eqs. (\ref{eqc1}), (\ref{eqc2}), and (\ref{eqb}). This indicates
that the pseudomode is just the discrete quasimode in the situation
of Lorentzian spectral density.

\section*{ACKNOWLEDGMENTS}

This work is supported by the Key Project of the National Natural
Science Foundation of China (Grant No. 60837004), and the National
Hi-Tech Program of China (863 Program). The authors appreciate Han
Pu for his reviewing  the manuscript.

\end{document}